\documentclass{cflowproc}

\usepackage{natbib}

\begin{document}


\shortauthors{P. M. Motl et al.}     
\shorttitle{Star Formation in Galaxy Clusters} 

\title{The Impact of Star Formation on Cool Core Galaxy Clusters}   

\author{Patrick M. Motl,\affilmark{1}   
Jack O. Burns, \affilmark{1}                
Michael L. Norman \affilmark{2}
and Greg L. Bryan \affilmark{3}}

\affil{1}{University of Colorado, Center for Astrophysics and Space Astronomy, Boulder, CO 80309}   
\and                                
\affil{2}{University of California San Diego, Center for Astrophysics and Space Sciences, 9500 Gilman Drive, La Jolla, CA 92093}
\and
\affil{3}{University of Oxford, Astrophysics, Keble Road, Oxford OX1 3RH}


\begin{abstract}
 We present results from recent simulations of the formation and evolution of clusters of galaxies 
in a $\Lambda$CDM cosmology. These simulations contain our most physically complete input physics 
to date including radiative cooling, star formation that transforms rapidly cooling material into 
aggregate star particles and we also model the thermal feedback from resulting supernovae in the 
star particles. We use an adaptive mesh refinement (AMR) Eulerian hydrodynamics scheme to obtain 
very high spatial resolution ($\approx 2 \; kpc$) in a computational volume 256 Mpc on a side with 
mass resolution for dark matter and star particles of $\approx 10^{8} \; M_{\odot}$. We examine in 
detail the appearance and evolution of the core region of our simulated clusters. 
\end{abstract}


\section{Description}
\label{motl:description}

We explore the role that star formation plays
in shaping the appearance of gas in clusters of galaxies.  If
only radiative cooling is allowed to operate, nearly
all halos in our simulations develop cores of cool, dense material
at their center.  Star formation provides a natural sink for this
cool material and additionally, the feedback of energy from supernovae
may impact the energy budget of some clusters and groups.  In this
poster, we describe our simulation efforts with star formation and
demonstrate its impact by examining (1) the thermodynamic phase of material
in representative simulations, (2) the run of
entropy in simulated clusters and (3) the temperature
profiles from samples of simulated clusters.  We also present preliminary
results from simulations where star formation is halted at a specific
redshift.  These simulations with truncated star formation yield promising
results for the formation of realistic cool
cores in clusters of galaxies.  Also, there is a sequence of movies available
that illustrate the evolution of one massive cluster at high resolution.

\section{Simulations}
\label{motl:simulations}

We have constructed samples of simulated galaxy
clusters using a sophisticated, Eulerian adaptive mesh refinement
cosmology code that incorporate successively more sophisticated input
physics.  The simulations evolve dark matter particles and utilize
the piecewise parabolic method to evolve the baryonic component.
Each simulation sample contains the same
clusters evolved with a specific set of physical assumptions at modest resolution
(15.6 kpc).
In our baseline adiabatic sample our simulations
trace the gravitational heating as clusters assemble.  At the
next level of complexity, we include radiative cooling by the cluster
gas.  We use a tabulated cooling curve assuming a metal abundance of
0.3 solar.  We have also generated cluster samples using a star formation
algorithm similar to the scheme developed by Cen \& Ostriker (1992)
that transforms rapidly cooling, collapsing gas into collisionless
``star'' particles.
For the star formation samples, we consider both the case of
star formation only and the case where the stars deposit
thermal energy back into the neighboring gas.

Our star formation algorithm sweeps through the finest resolution grids every
timestep and searches for cells with the following properties:
\begin{itemize}
  \item Resides in a region of overdensity exceeding a threshold, $\delta$
  \item Convergence of flow in the cell ($\mathbf{\nabla} \cdot \mathbf{v} < 0$)
  \item The local cooling time is shorter than the local free fall time
  \item The mass of baryons in the cell exceeds the Jean's mass.
\end{itemize}
If a cell that matches these conditions has enough material to exceed a minimum
star particle mass (which is introduced to limit the number of star particles that
must be evolved) then a new star is created with a mass $m_{\star} = \eta \; m_{\mathrm{baryon}} \;
\Delta t / t_{\mathrm{dynamical}}$, where $\Delta t$ is the simulation timestep and $\eta$ is
an efficiency parameter.
If the minimum mass threshold is not met in this timestep, the code will track the
frustrated attempt to form a star and the threshold can be exceeded incrementally.
Once formed, the star begins to both heat the surrounding fluid to model the feedback
from prompt supernovae and the star pollutes the fluid with a passive metallicity
tracer field.  The total energy deposited is scaled to the rest-mass energy of the
star particle with an efficiency  parameter, $\epsilon$.

Given the complexity of the star formation algorithm, we have run
numerous simulations to explore the parameter space.  Variations in the overdensity threshold,
$10 < \delta < 1 \times 10^{4}$; feedback strength, $0 < \epsilon < 1 \times 10^{-4}$;
star formation efficiency, $0.01 < \eta < 1$; and minimum
mass of the star particles (ranging from $10^{6} M_{\odot}$ to $10^{9} M_{\odot}$)
do not significantly impact the properties of the baryons in clusters of galaxies.
Roughly, the strength of feedback controls the amount of mass that is converted into
stars and to a lesser degree, the overdensity threshold dictates when star
formation begins in the simulation.   Similarly, the efficiency of star formation will
control the amount of mass converted into stars but to a lesser degree than the feedback strength.
\textbf{On the scale of clusters however, the primary
role of star formation is to allow rapidly cooling gas to leave the fluid.}

All simulations presented here
were computed assuming a $\Lambda$CDM cosmological model
with the following parameters: $\Omega_{\mathrm{b}} = 0.026$, $\Omega_{\mathrm{m}} = 0.3$,
$\Omega_{\Lambda} = 0.7$, $\mathrm{h} = 0.7$, and $\sigma_{8} = 0.928$.
Each physics sample contains approximately 75 clusters in the mass
range from $4 \times 10^{14}$ M$_{\odot}$ to $2 \times 10^{15}$
M$_{\odot}$ at the present epoch.

\section{Baryon Distribution}
\label{motl:baryons}

Radiative cooling allows the baryonic fluid to lose energy, causing the
material to lose pressure support and condense.  With the addition of star formation,
these cool condensations can be removed from the fluid.  However, the gross thermodynamic
state of the baryons is dictated by the structure formation process and only relatively
small regions in the simulation have cooling times that are short on  a cosmological
timescale.  Furthermore,  it s only in these regions where star formation can act.
As can be seen in Figure 1, the overall thermodynamic state of baryons in the simulations
does not depend strongly on the input physics beyond the hierarchical collapse and shock
heating present in the adiabatic limit.  In highly  overdense regions,
such as the cores of clusters of galaxies, additional physics such as star formation
play a significant role.

\begin{figure}
      \begin{tabular}{c}
         \textbf{Adiabatic} \\
         \includegraphics[scale=0.4]{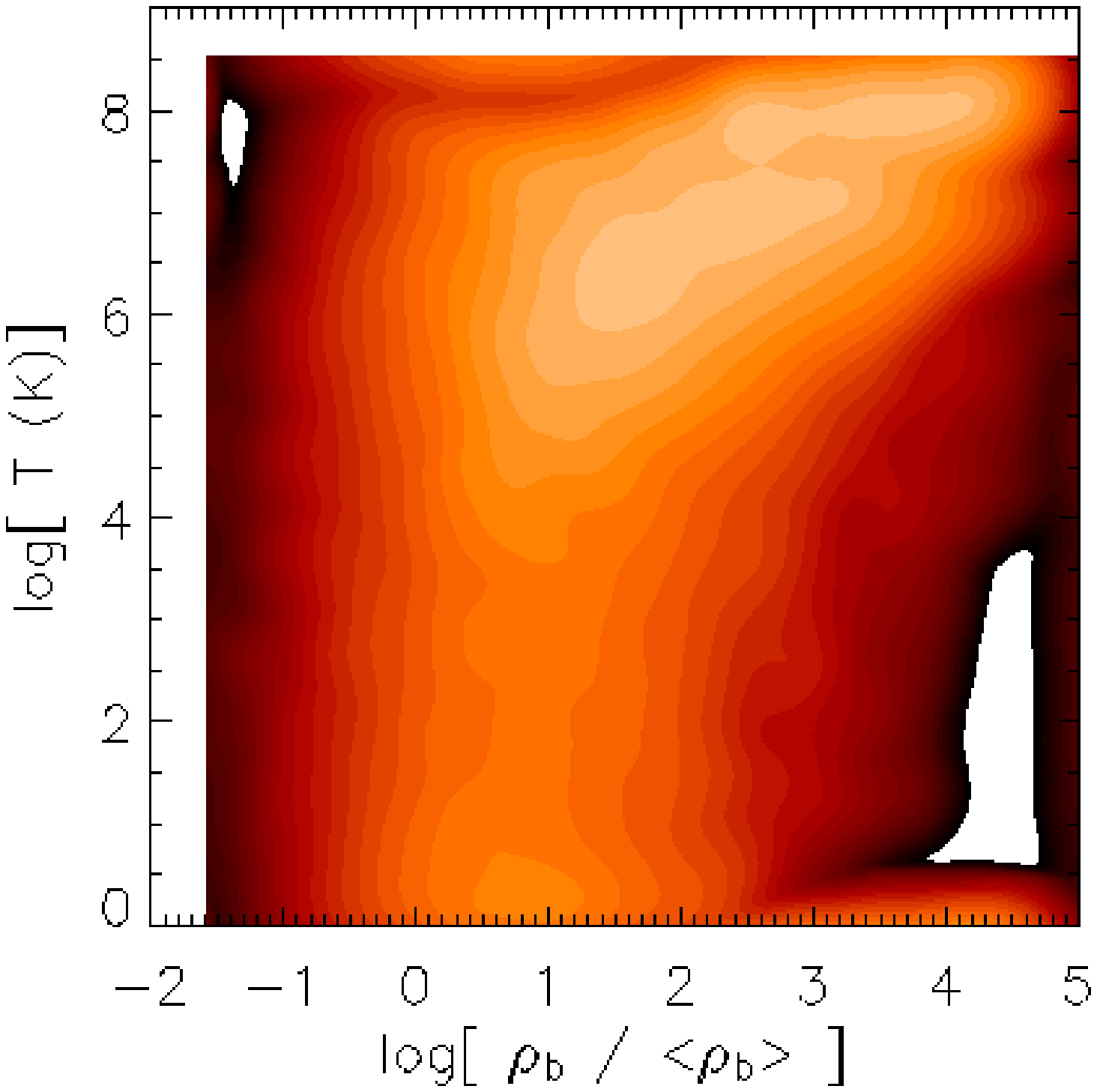} \\
         \textbf{Radiative Cooling} \\
         \includegraphics[scale=0.4]{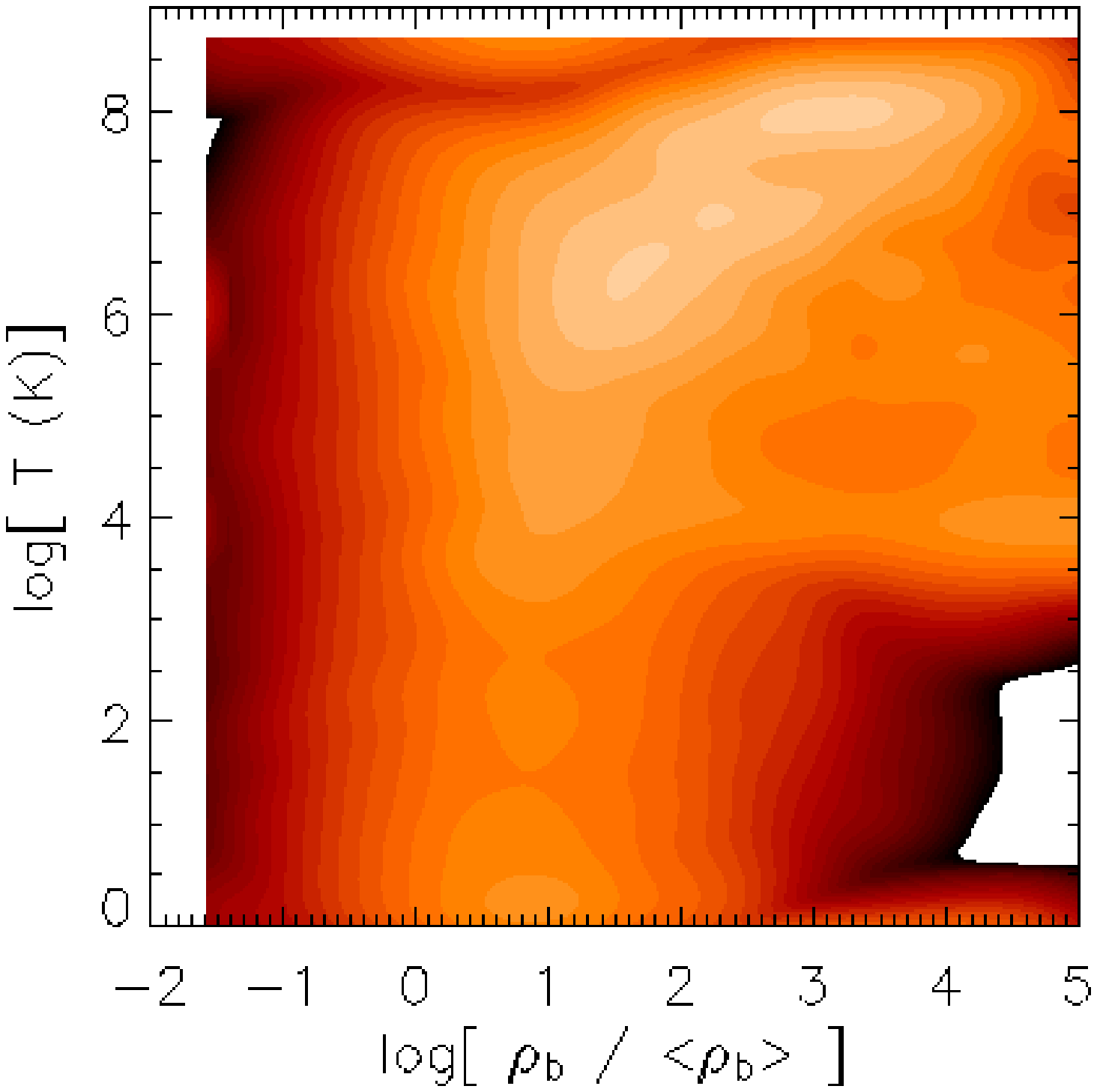} \\
         \textbf{Star Formation + Feedback} \\
         \includegraphics[scale=0.4]{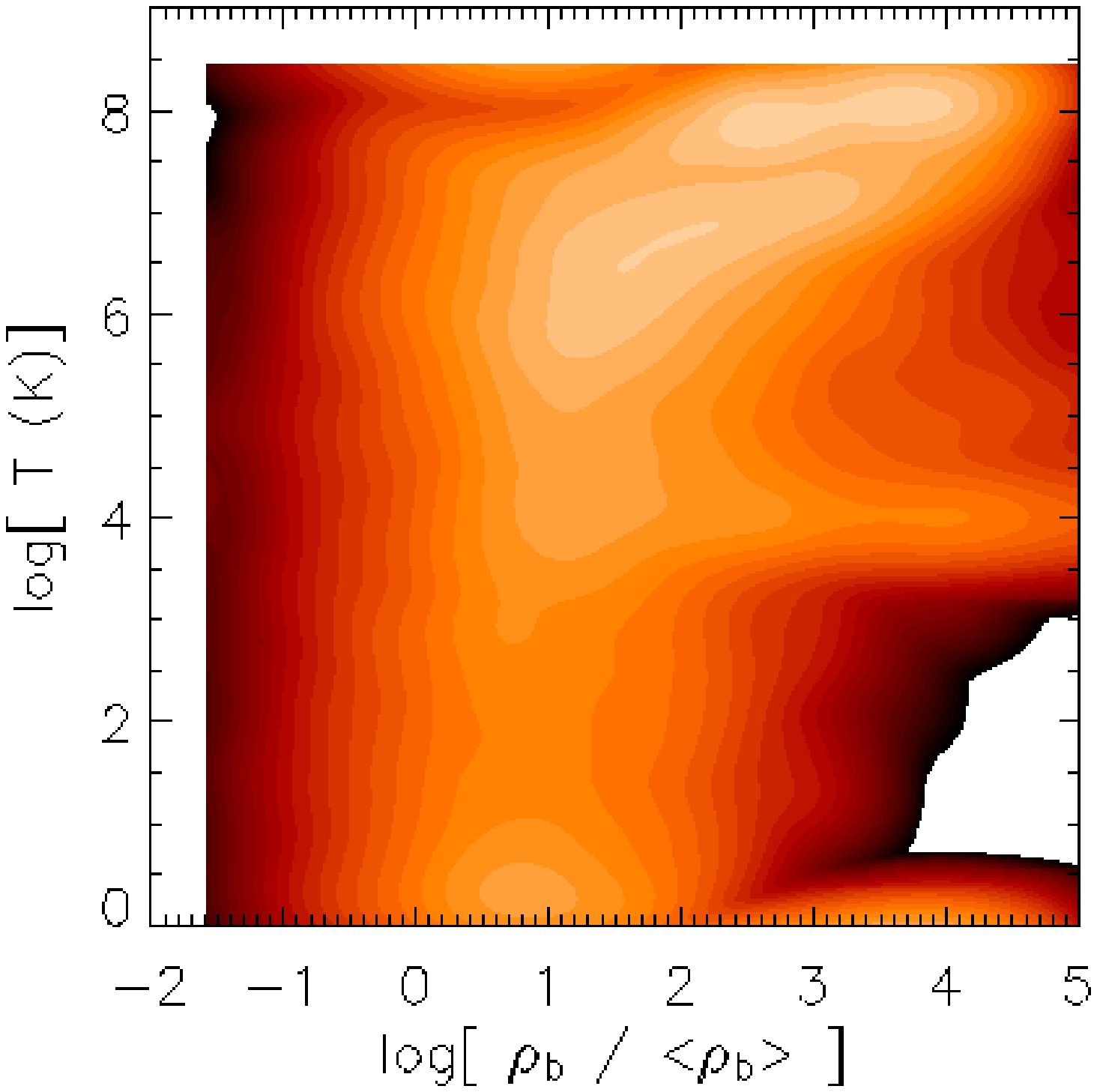} \\
      \end{tabular}
   \figcaption{The thermodynamic phase of all gas cells in the refined region for three
    different sets of input physics at the present epoch.  The contours are
    in the fraction of total mass in the volume at the indicated
    temperature and density.  Note that the inclusion of star formation and feedback
    does not significantly alter the distribution of baryons.}
\end{figure}

\section{The Entropy Floor}
\label{motl:entropy}

The process of galaxy formation in the distant past is believed to
have left a signature on the present day appearance of clusters and groups
in the form of a universal entropy floor (Ponman \textit{et al.} 1999).
Gas at a lower entropy than this universal floor value should not appear in clusters
as it would have been either heated to higher entropies by non-gravitational
processes or either cooled to invisibility or collapsed into stars due to its
short cooling time.  In Figure 2 we show profiles of the fluid's entropy - normalized
by the average cluster temperature - for 50 clusters evolved with star
formation and a realistic amount of thermal feedback.
The profiles have been color coded by the average cluster temperature
as indicated in the figure caption.
Most clusters have normalized core
entropies at $ \approx 100 \; cm^{2}$
and therefore, these systems approximately obey self-similar scaling ($s \propto T$)
in their cores.
We do note that the variation of core entropy increases significantly for the cooler
systems.
A plot of entropy profiles for the sample without supernova feedback, meaning that
the gas can collapse to form stars but these stars do not heat the remaining fluid,
is very similar to that shown in Figure 2.  This  indicates that supernova feedback does not
play a significant role in heating the cluster material, rather it is the case that
fluid with entropy lower than the floor value is simply removed from the system.
A puzzle remains however as other forms of non-gravitational heating, such as feedback
from AGN, would likely raise the core entropy beyond that observed. Yet, some physical
mechanism must be acting to break self-similarity if an entropy scaling relation
of the form $s \left(0.1 r_{\mathrm{virial}}\right) \propto T^{0.65}$ as measured by Ponman \textit{et al.}
(2003) is correct.

\begin{figure}
   \begin{center}
      \rotatebox{90}{\includegraphics[scale=0.35]{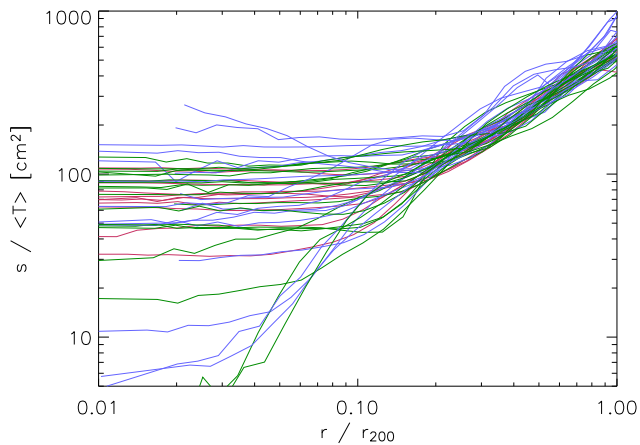}}
   \end{center}
   \figcaption{The entropy ($s \equiv T / n_{e}^{2/3}$) scaled by the mean cluster temperature
   as a function of normalized radius for 50 clusters in the continuous star formation +
   feedback sample.  The profiles are color-coded by the average temperature of the cluster
   with red denoting clusters with $5 \; keV < T_{\mathrm{avg}} < 10 \; keV$, green for clusters with
   $3.6 \; keV < T_{\mathrm{avg}} < 5 \; keV$ and finally blue for clusters with $2.4 \;  keV < T_{\mathrm{avg}} < 3.6 \; keV$.
   The average cluster temperature is computed from the emission-weighted temperature within
   0.3 virial radii for an overdensity of 200.}
\end{figure}

\section{Are There Cool Cores with Continuous Star Formation?}
\label{motl:cool_cores}

A crucial test of our simulations is whether they produce clusters with temperature
structure similar to that in systems observed with \textit{Chandra} and \textit{XMM}.  As a simple
first step toward making this comparison, we examine the temperature profiles from our sample
clusters.  In Figure 3, we show the average profiles from 50 clusters evolved with adiabatic
physics, radiative cooling, star formation and finally star formation with feedback.  For both star
formation samples, no cluster exhibits a cool core at the present epoch whereas a majority of observed clusters
have cool cores.  Instead, the individual cluster temperature
profiles are flat or slightly rising toward the cluster center of mass.  We have confirmed this result with
simulations of individual clusters at higher resolution (up to 1 kpc) and with different input parameters for the
star formation algorithm.

\begin{figure}
   \begin{center}
      \rotatebox{90}{\includegraphics[scale=0.35]{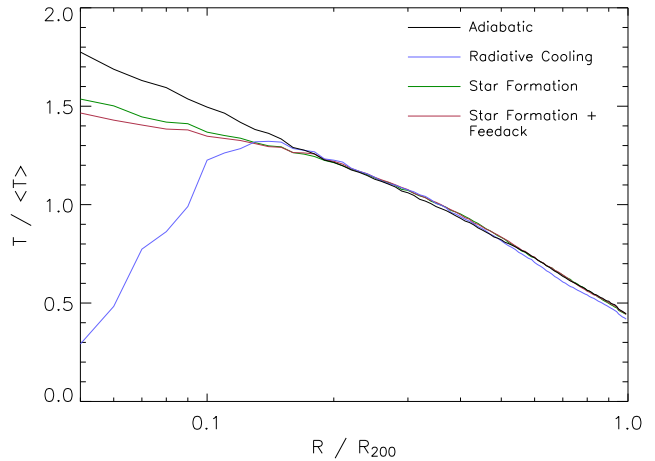}}
   \end{center}
   \figcaption{The average temperature profiles derived from the projected, emission-weighted temperature
   field for our four physics samples plotted against the normalized radius.  The average temperature has been
   calculated within half a virial radius of the cluster center of mass.}
\end{figure}

\section{Truncated Star Formation}
\label{motl:truncz}

We have recently studied a simple extension to our standard star formation
prescription that is meant to model the observed decay in global star formation rate
at a redshift between 1 and 2.  As a first approximation to enforcing this
constraint on our simulations, we have simply turned off star formation at
a given redshift and allowed the simulation to complete with losses due
to radiative cooling and, for consistency, with feedback from the star
particles that have already formed.  As can be seen in Figure 4, the epoch
where star formation is truncated ($z_{truncation}$) strongly influences the
evolution of the cluster if it occurs in a narrow range between redshifts of
1.5 and 2.
While these results are both preliminary and are calculated in a rather
crude approximation, it is important to note that this approach represents
the only means we have found to date for creating a realistic cool-core cluster
with star formation,  meaning a cluster with a core of material at about 1-2 keV.

\begin{figure*}
  \begin{center}
   \begin{tabular}{cccc}
        \multicolumn{4}{c}{\textbf{Radiative Cooling}} \\
        \rotatebox{90}{\includegraphics[scale=0.4]{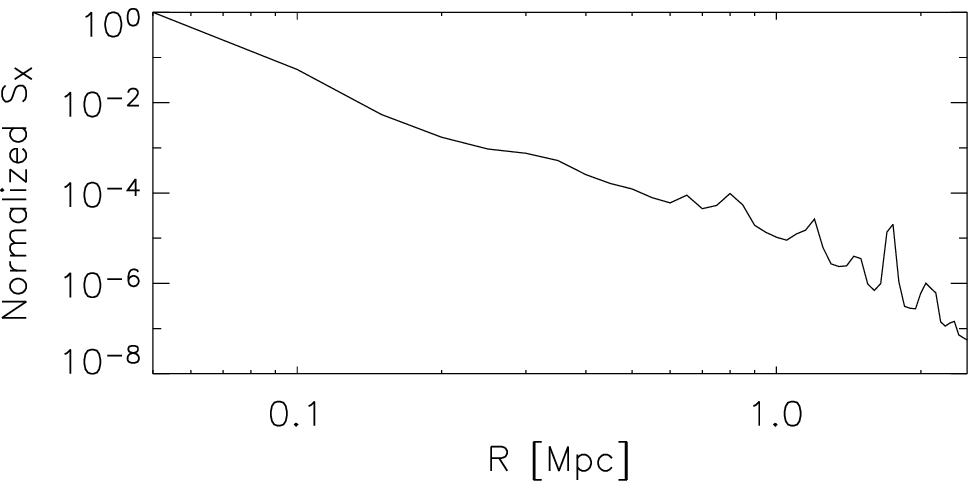}} &
        \includegraphics[scale=0.4]{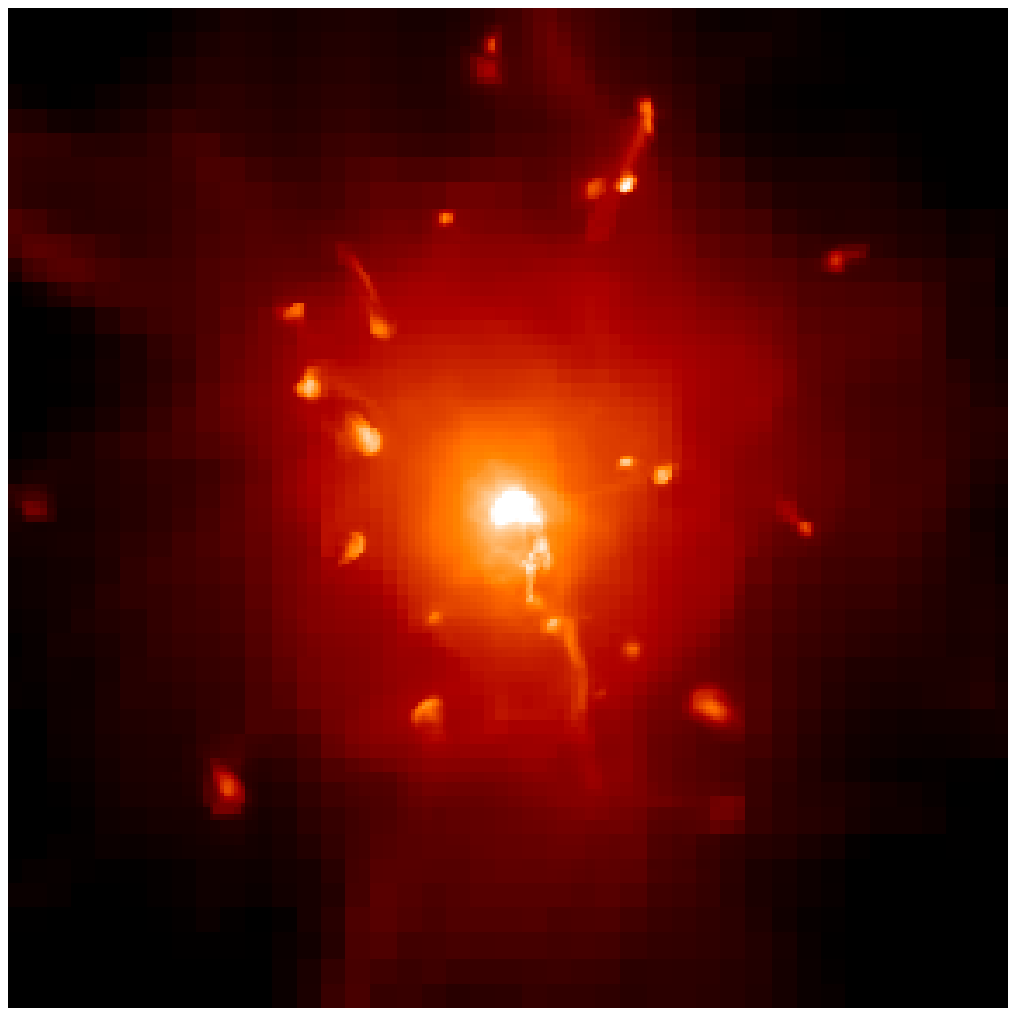} &
        \includegraphics[scale=0.4]{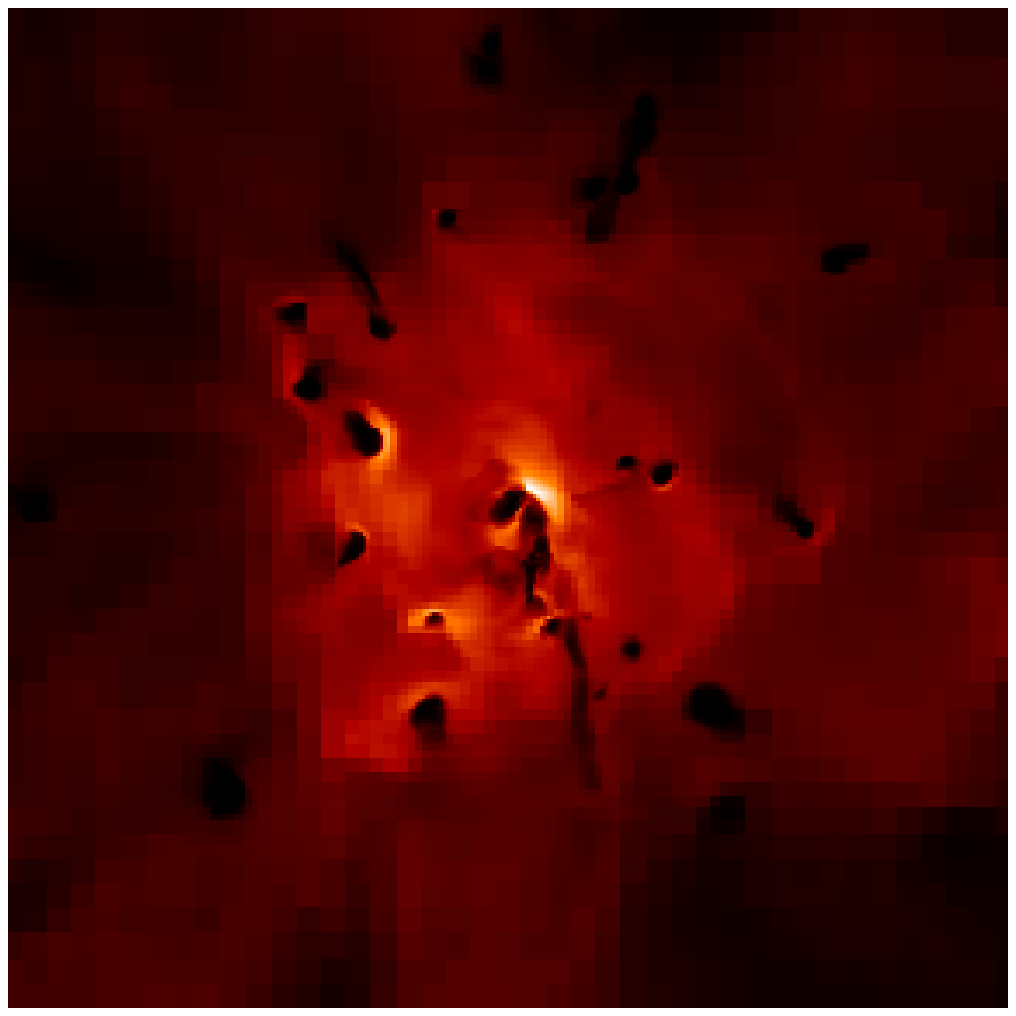} &
        \rotatebox[x=0.6in,y=1.0in]{270}{\includegraphics[scale=0.4]{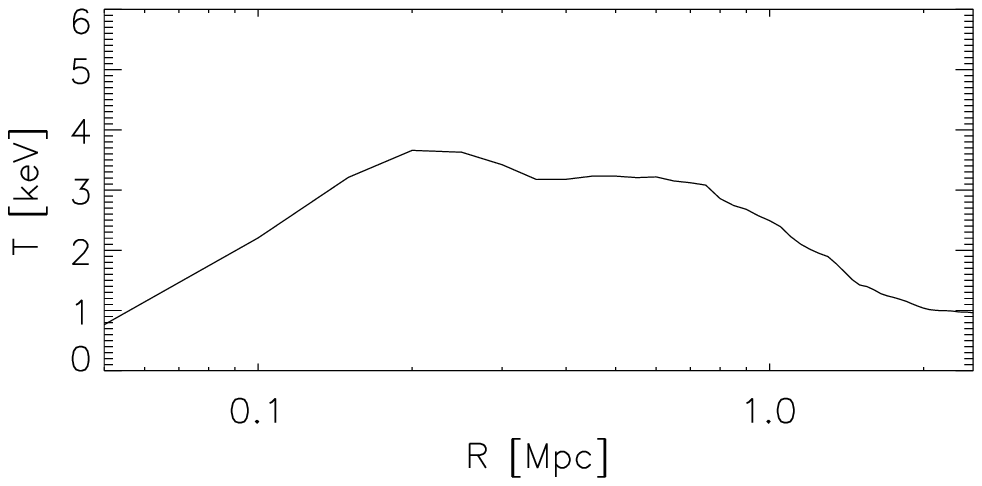}} \\
        \multicolumn{4}{c}{ \textbf{$z_{truncation} = 2$}} \\
        \rotatebox{90}{\includegraphics[scale=0.4]{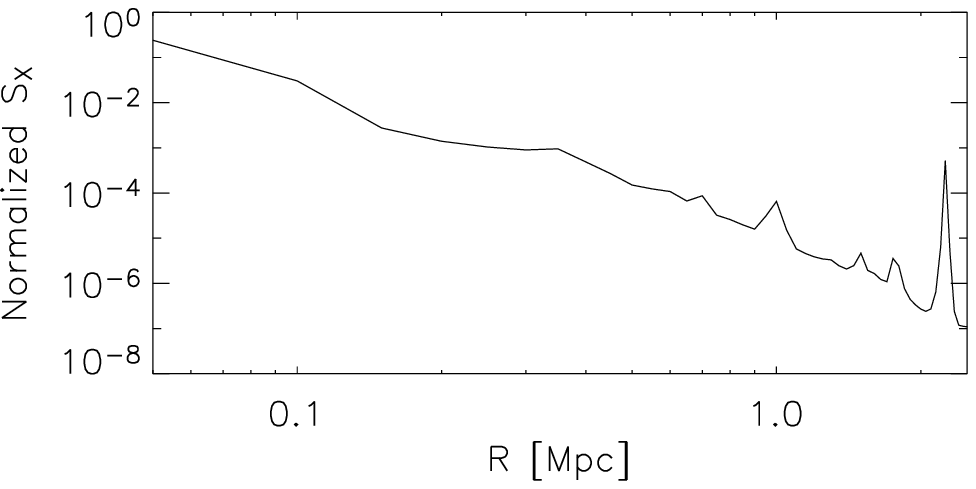}} &
        \includegraphics[scale=0.4]{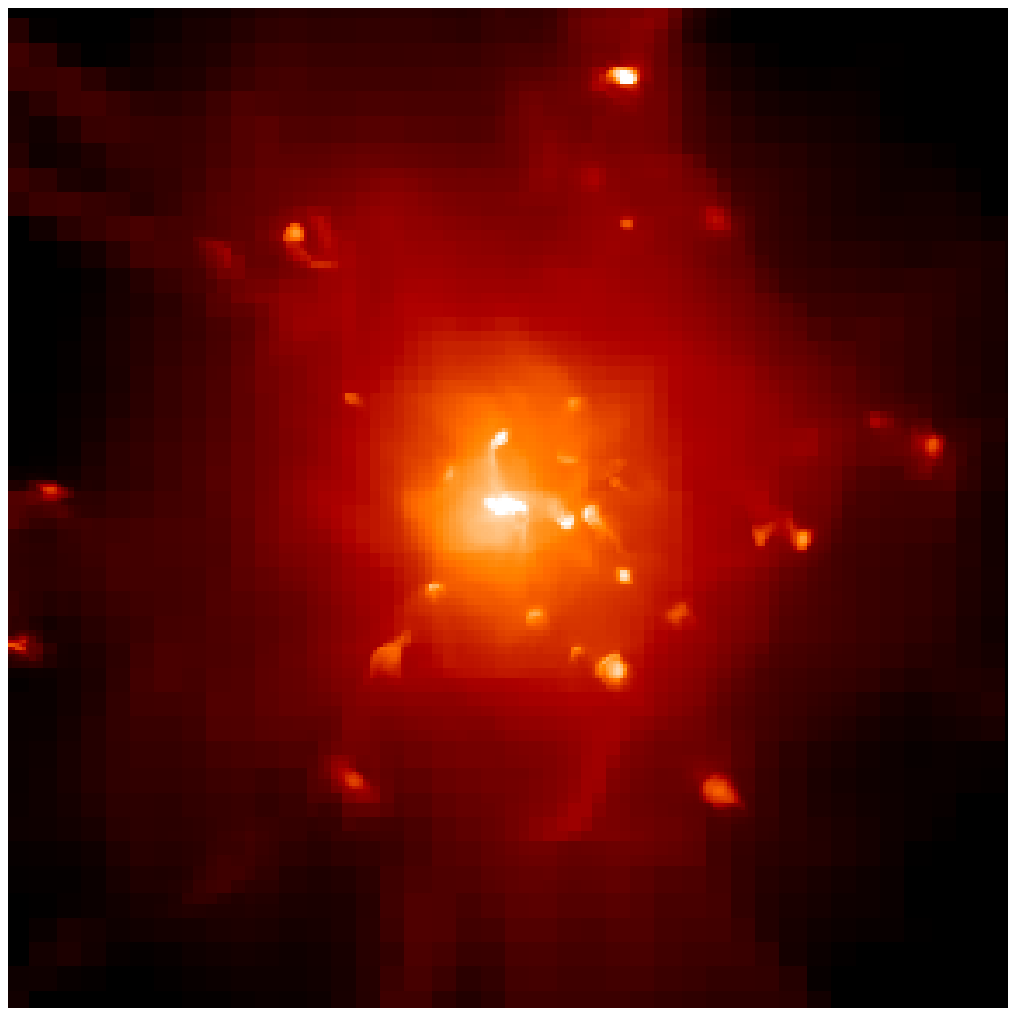} &
        \includegraphics[scale=0.4]{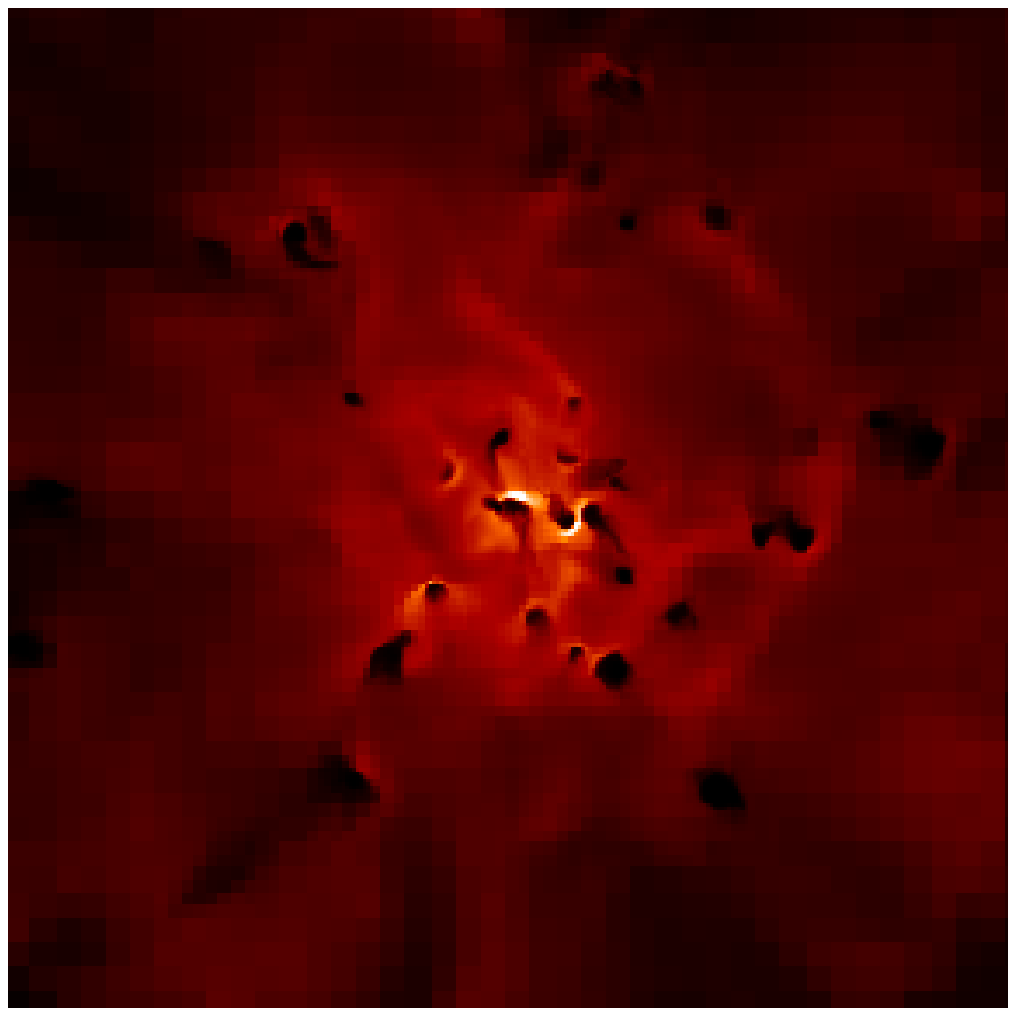} &
        \rotatebox[x=0.6in,y=1.0in]{270}{\includegraphics[scale=0.4]{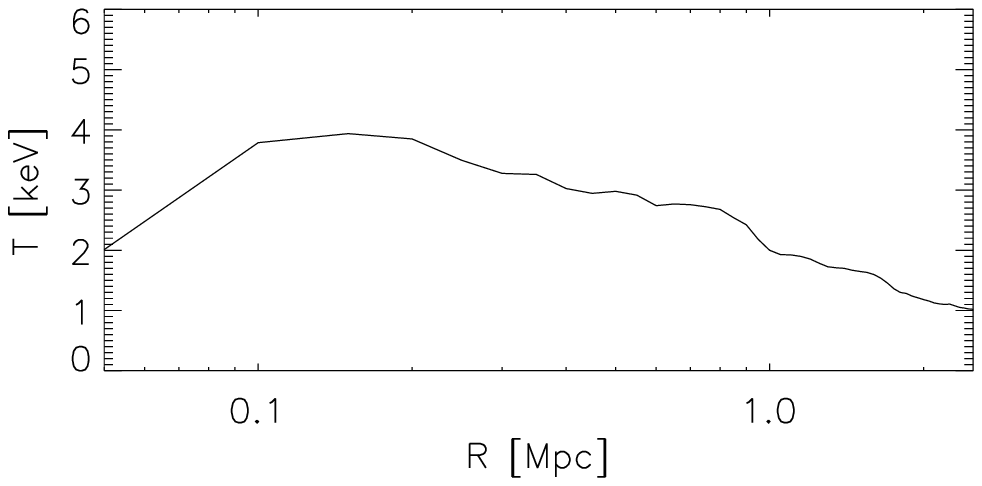}} \\
        \multicolumn{4}{c}{ \textbf{$z_{truncation} = 1.5$}} \\
        \rotatebox{90}{\includegraphics[scale=0.4]{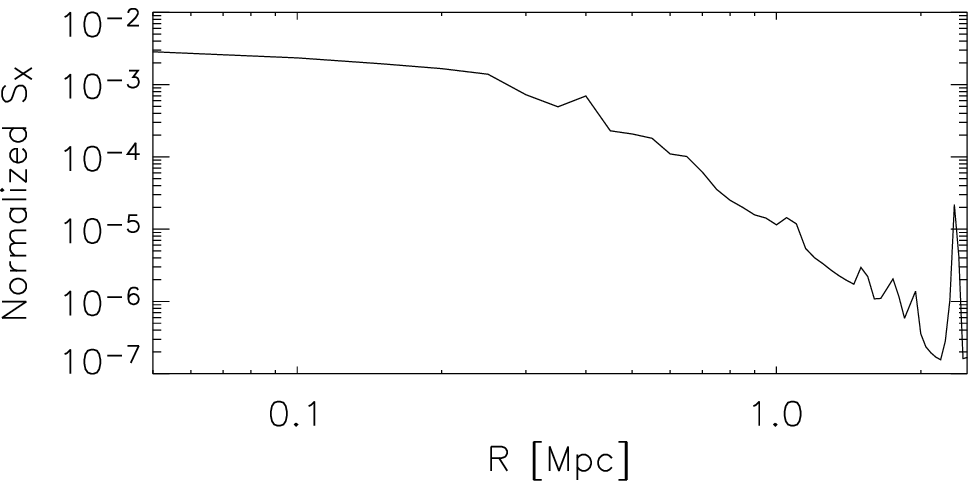}} &
        \includegraphics[scale=0.4]{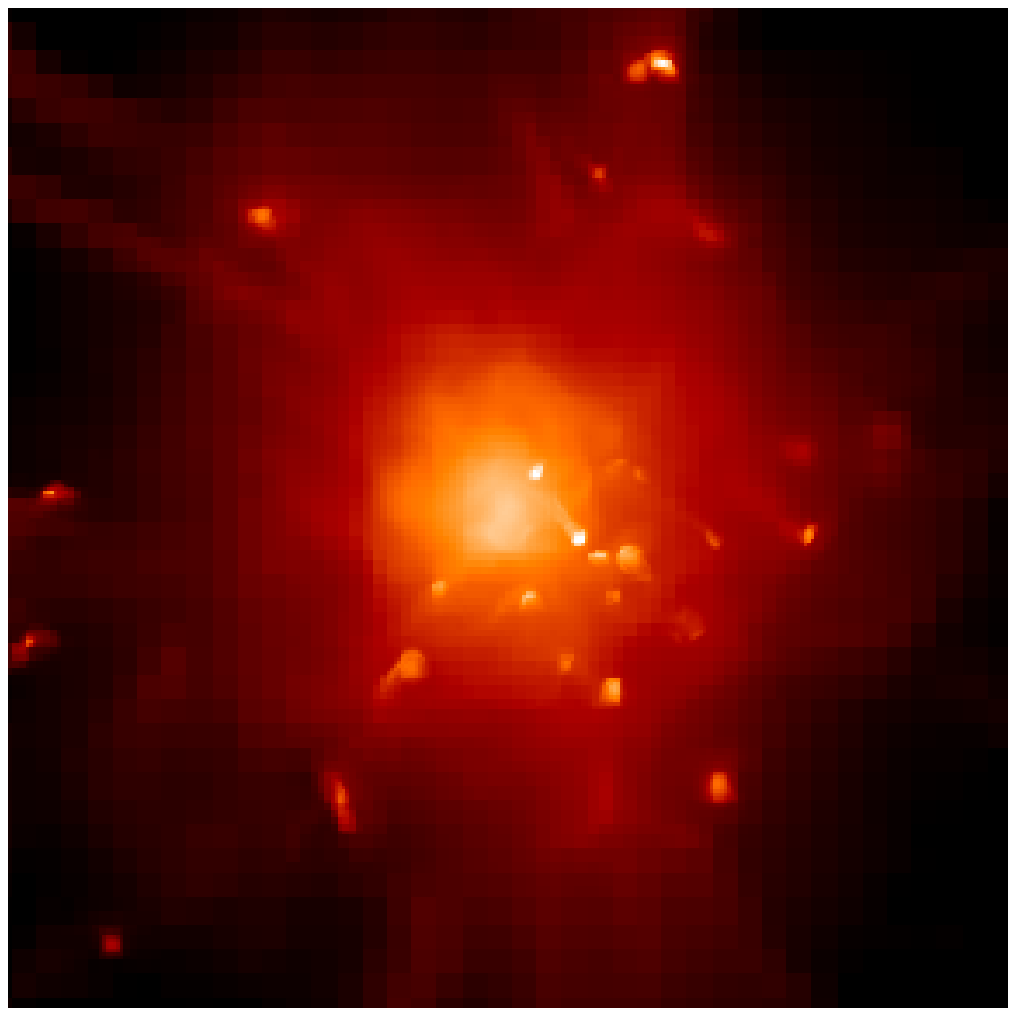} &
        \includegraphics[scale=0.4]{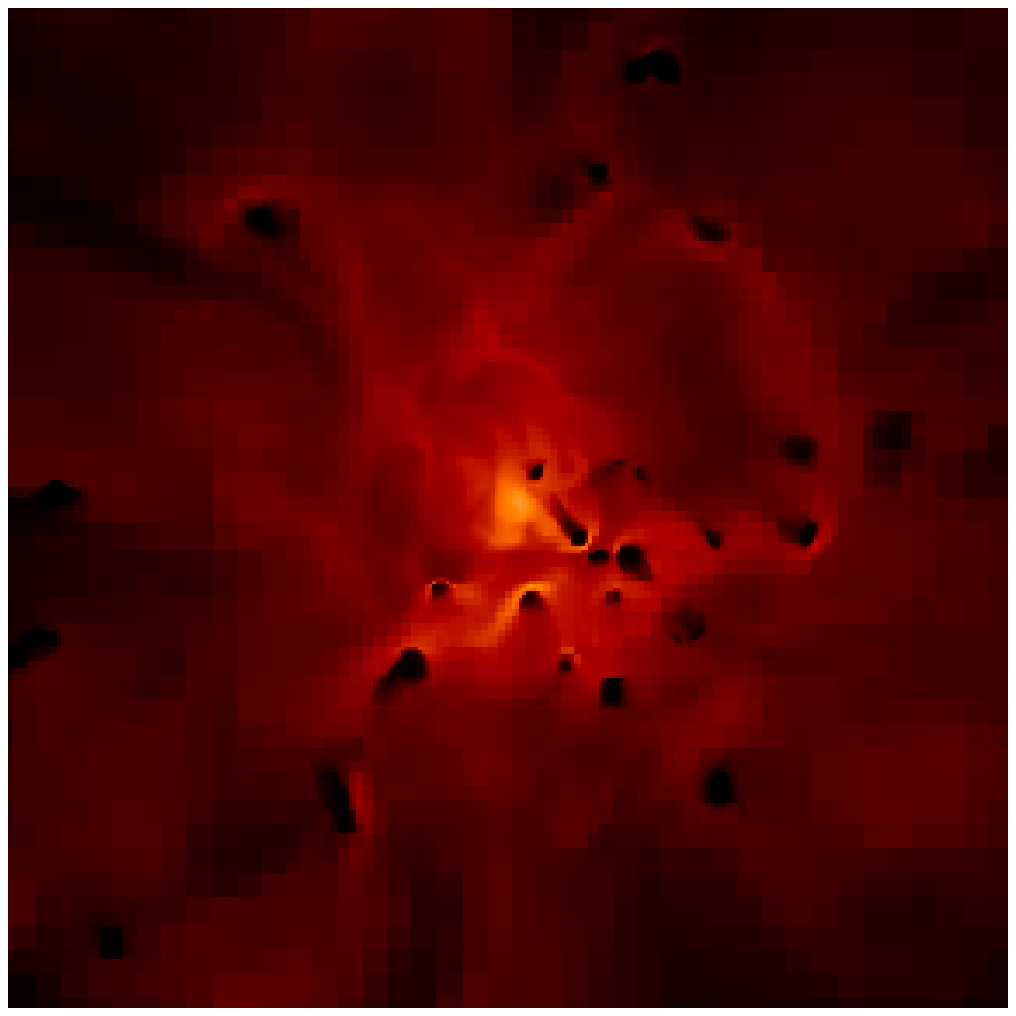} &
        \rotatebox[x=0.6in,y=1.0in]{270}{\includegraphics[scale=0.4]{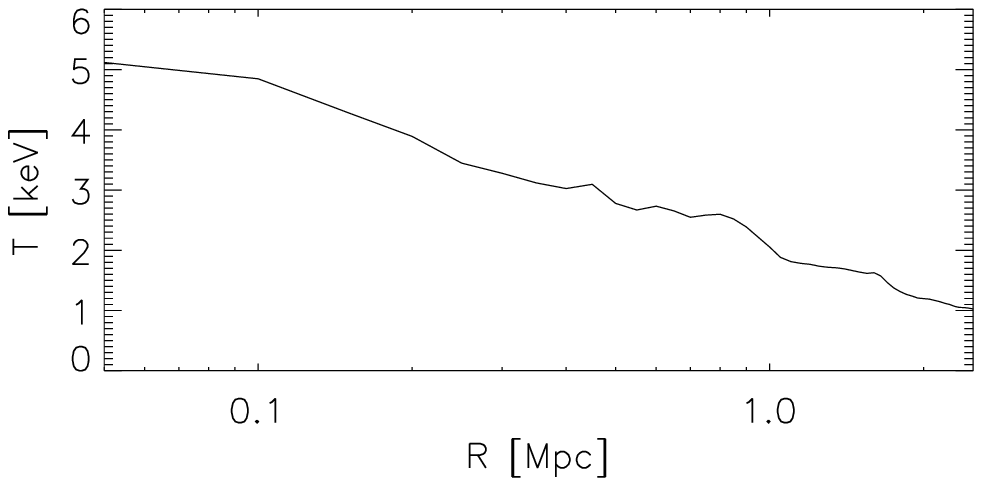}} \\
        \multicolumn{4}{c}{ \textbf{$z_{truncation} = 0$}} \\
        \rotatebox{90}{\includegraphics[scale=0.4]{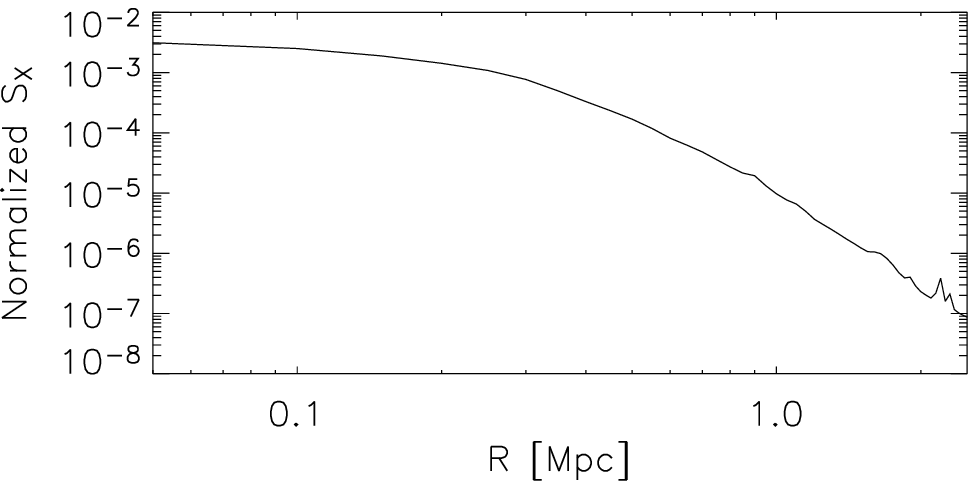}} &
        \includegraphics[scale=0.4]{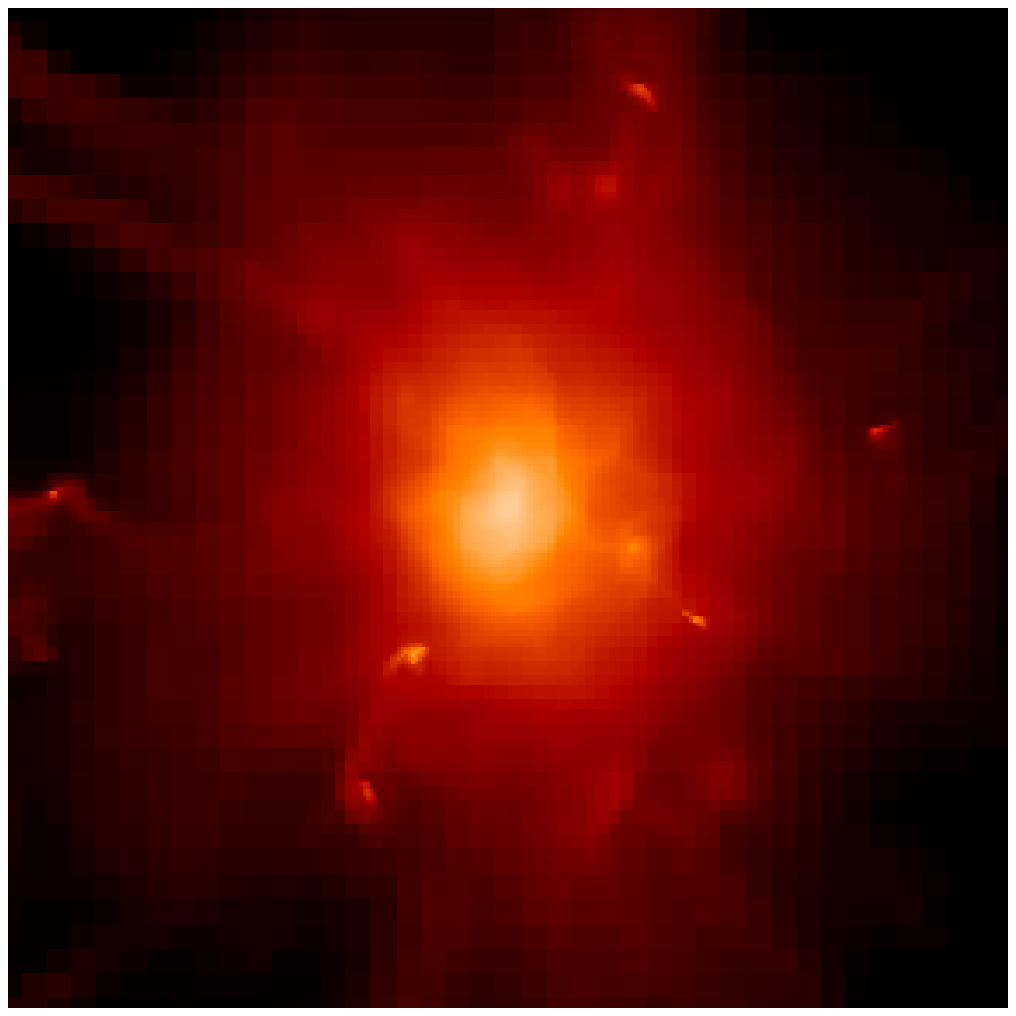} &
        \includegraphics[scale=0.4]{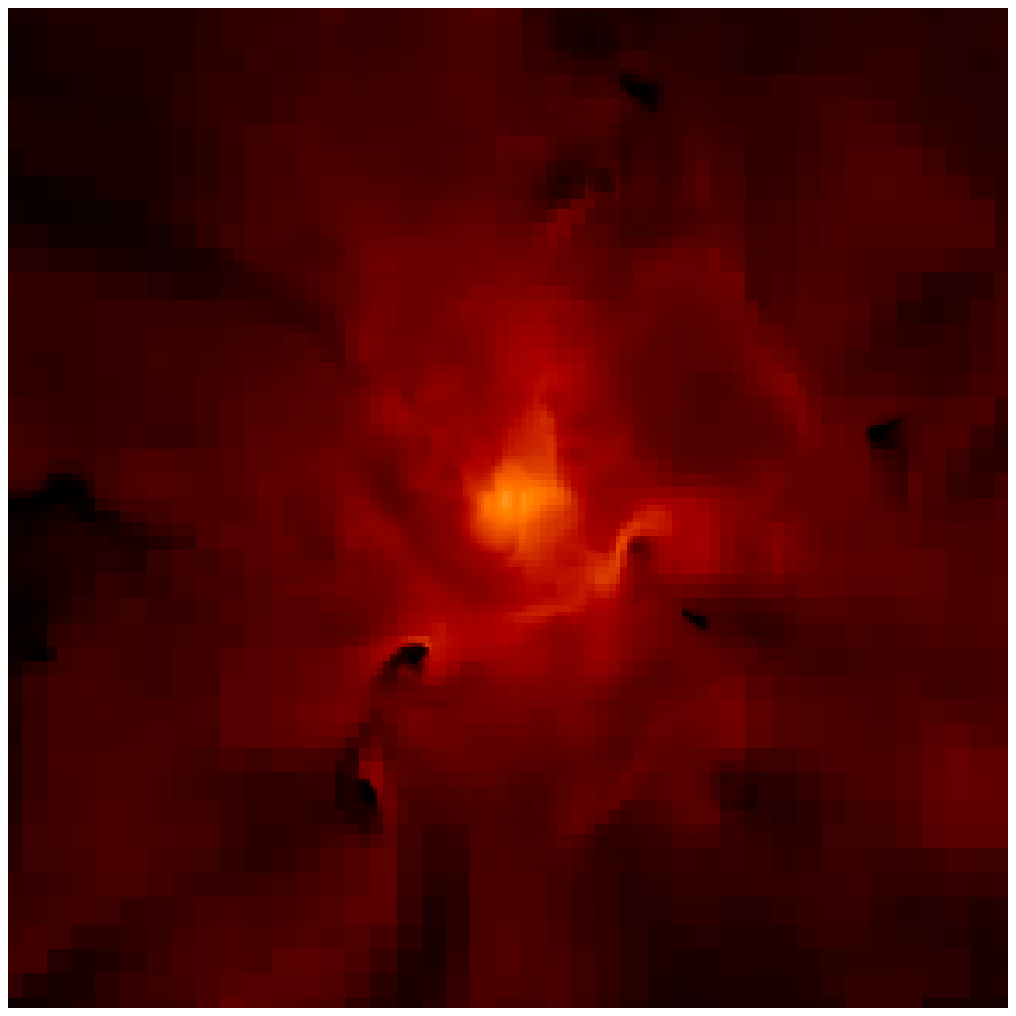} &
        \rotatebox[x=0.6in,y=1.0in]{270}{\includegraphics[scale=0.4]{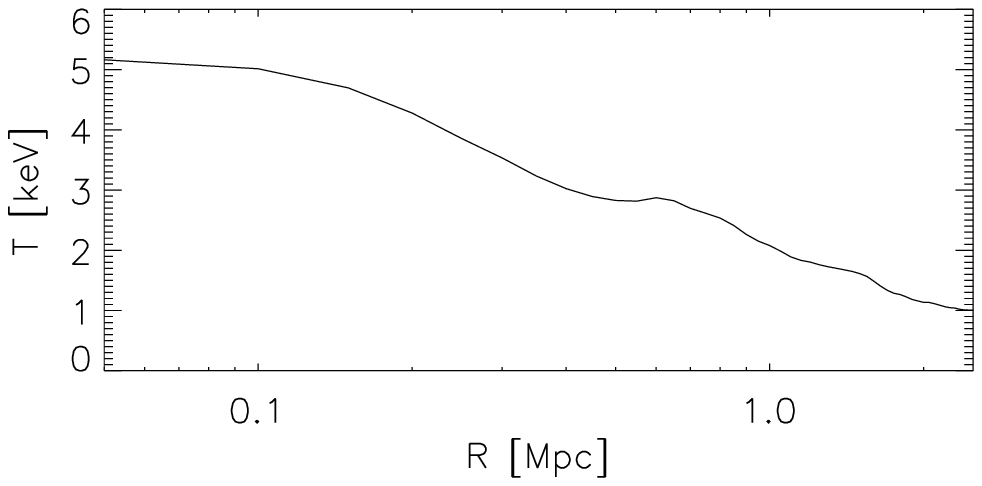}} \\
        \multicolumn{4}{c}{ \textbf{Adiabatic}} \\
        \rotatebox{90}{\includegraphics[scale=0.4]{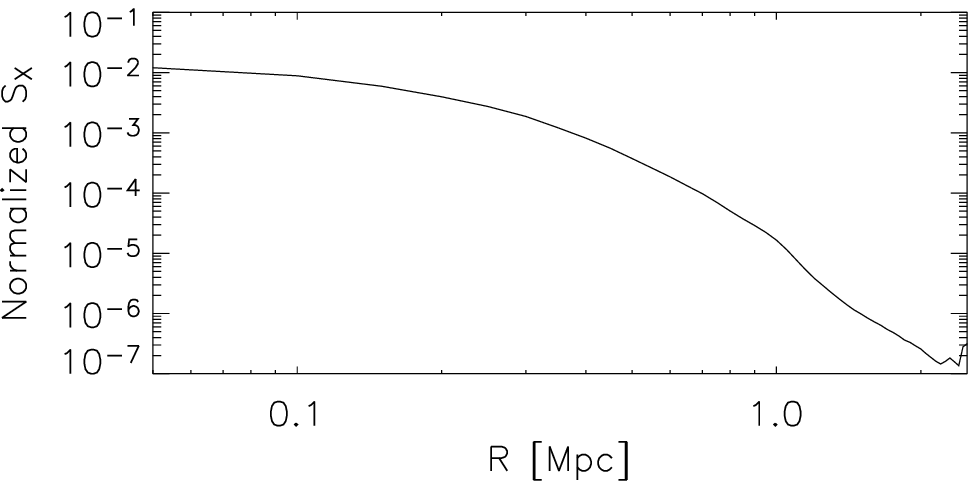}} &
        \includegraphics[scale=0.4]{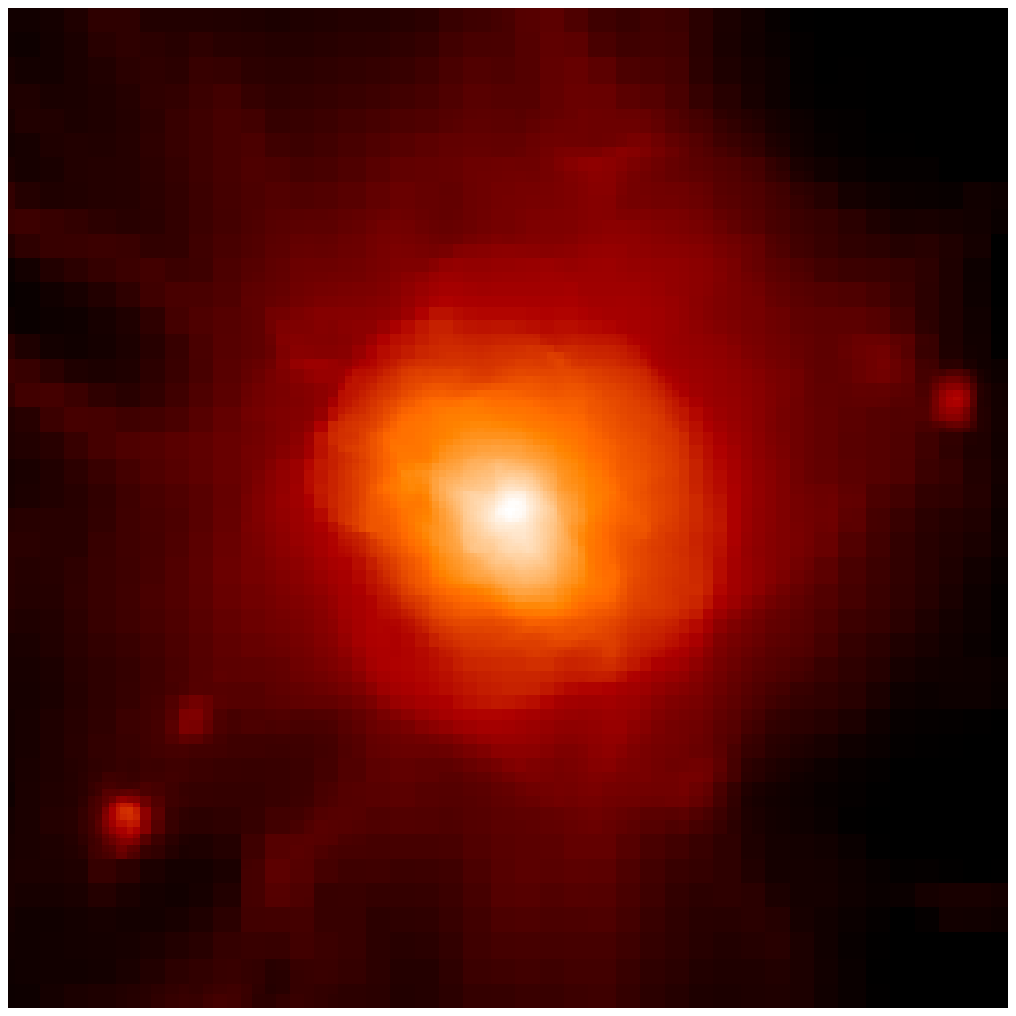} &
        \includegraphics[scale=0.4]{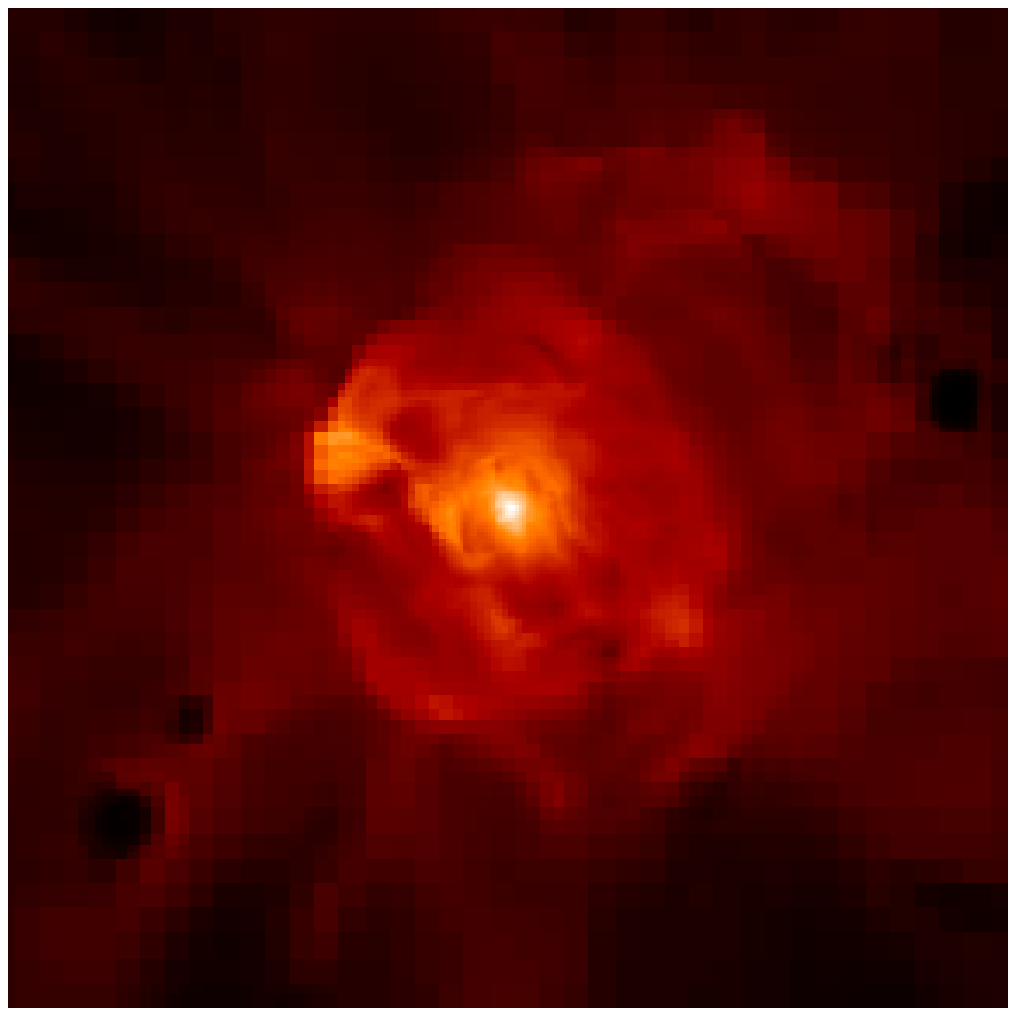} &
        \rotatebox[x=0.6in,y=1.0in]{270}{\includegraphics[scale=0.4]{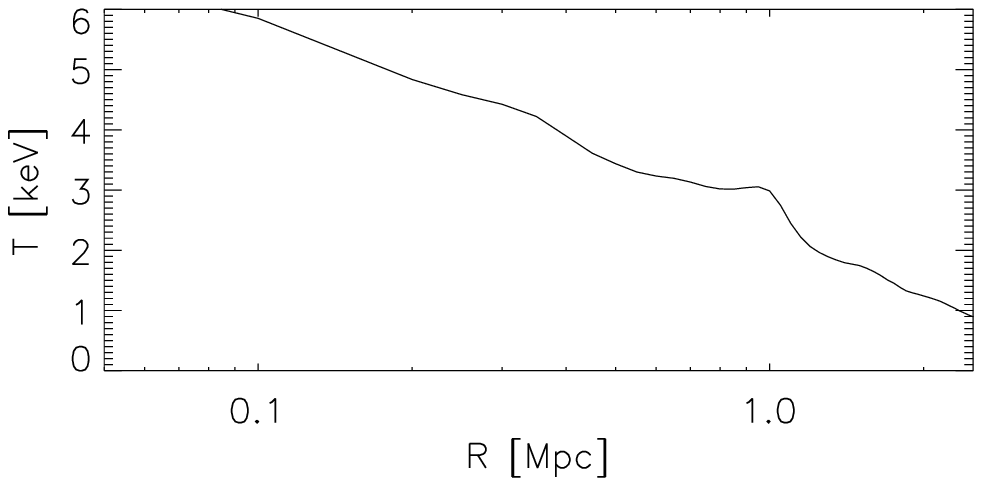}} \\
   \end{tabular}
   \end{center}
   \figcaption{Images of X-ray surface brightness and projected, emission-weighted
temperature maps at the present epoch for the same cluster run in the adiabatic and radiative cooling limits
as well as with star formation + feedback for a range of truncation values.  Each image is 5 Mpc on a side.
Profiles of the normalized
X-ray surface brightness and projected temperature are shown in the far left and right columns.   If star
formation is truncated at redshifts higher than 2, the cluster forms a cool core.  If, however, star formation
continues past a redshift of 1.5 no gas remains with a sufficiently short cooling time to form a
cool core at the present epoch.}
\end{figure*}

\section{Movie Description}
\label{motl:movie}

We have prepared a series of animations for one particular cluster evolved with star formation
and supernova feedback.  The animations are available in mpeg format at the following URL,
http://casa.colorado.edu/$\sim$motl/research.
The simulation was run with 10 levels of refinement yielding a peak spatial resolution
of slightly less than $2 \; kpc$ and with the following parameters for the star formation algorithm (see the discussion of
the simulation technique above for a description of these parameters): $\delta = 100$, $\eta = 0.1$,
$\epsilon = 4.1 \times 10^{-6}$, and a minimum star particle mass of $1 \times 10^{7} M_{\odot}$.  In addition,
we assume that a given star particle will lose one quarter of its initial mass during supernova feedback and
that the ejecta is polluted with metals at solar abundance (2\% by mass).
The simulation evolves a box approximately 40 Mpc on a side with adaptive mesh refinement
within a volume 256 Mpc on a side.  In the movies, we show a much smaller region 8 Mpc on a side at the present
epoch that is centered about the final cluster center of mass.  Scalar quantities of interest for the cluster
at a redshift of zero are listed in the following table.  The virial overdensity is taken to be 200. Masses are
integrated out to the virial radius, as is the X-ray luminosity which is calculated for the 1 to 10 keV band.
$T_{\mathrm{avg}}$ is the average, emission-weighted temperature out to the virial radius while $T_{\mathrm{core}}$ is measured
over the central 50 kpc only.  The central cooling time is given as $t_{\mathrm{cool}}$ and $\sigma$ is the three-dimensional
average velocity dispersion.

\begin{center}
   \begin{tabular}{cc|cc}
      \hline
      $R_{\mathrm{virial}}$ & $2.6 \; Mpc$ & $L_{X}$ & $2.9 \times 10^{44} \;  erg  \; s^{-1}$ \\
      $M_{\mathrm{virial}}$ & $2.1 \times 10^{15} \; M_{\odot}$ & $T_{\mathrm{avg}}$ & $6.1 \;  keV$ \\
      $M_{\mathrm{dm}}$ & $2.0 \times 10^{15} \; M_{\odot}$ & $T_{\mathrm{core}}$ & $14  \; keV$ \\
      $M_{\mathrm{gas}}$ & $9.8 \times 10^{13} \;  M_{\odot}$ & $t_{\mathrm{cool}}$ & $8.7 \times 10^{10}  \; years$ \\
      $M_{\star}$& $2.1 \times 10^{13} \; M_{\odot}$ & $\sigma$ & $2,100  \; km  \; s^{-1}$ \\
   \end{tabular}
\end{center}

\section{Conclusion}

For our chosen star formation prescription with \textbf{continuous star formation},
no cluster has a cool core at the present epoch despite variation of the star formation parameters and
increasing the simulation resolution by a factor of 16.  However, in recent calculations,
\textbf{the truncation of star formation} at an epoch between redshifts of 1.5 and 2 produces a
realistic cool-core cluster.


\acknowledgements
This research was partially supported by grant TM3-4008A from NASA
The simulations presented in this poster were conducted on
the Origin2000 system at the National Center for Supercomputing Applications
at the University of Illinois, Urbana-Champaign through computer allocation
grant AST010014N.


\end{document}